\documentclass{article}
\usepackage{spconf,amsmath,graphicx}
\usepackage[utf8]{inputenc} 
\usepackage[T1]{fontenc}    
\usepackage{hyperref}       
\usepackage{url}            
\usepackage{booktabs}       
\usepackage{amsfonts}       
\usepackage{nicefrac}       
\usepackage{microtype}      
\usepackage{lipsum}
\usepackage{graphicx}
\usepackage{multirow}
\usepackage{makecell}
\usepackage[usenames,dvipsnames]{xcolor}
\usepackage{tikz}
\usepackage{pgfplots}
\usepackage[ruled,vlined]{algorithm2e}
\pgfplotsset{compat=1.17}

\usepackage{etoolbox}
\graphicspath{ {./images/} }

\def\ll{{\mathcal L}}

\makeatletter
\patchcmd{\@algocf@start}
  {-1.5em}
  {0pt}
  {}{}
\makeatother

\pgfplotsset{
    bar group size/.style 2 args={
        /pgf/bar shift={%
                -0.5*(#2*\pgfplotbarwidth + (#2-1)*\pgfkeysvalueof{/pgfplots/bar group skip})  + 
                (.5+#1)*\pgfplotbarwidth + #1*\pgfkeysvalueof{/pgfplots/bar group skip}},%
    },
    bar group skip/.initial=2pt,
    plot 0/.style={blue,fill=blue!30!white,mark=none},%
    plot 1/.style={red,fill=red!30!white,mark=none},%
    plot 2/.style={brown!60!black,fill=brown!30!white,mark=none},%
}

\title{Word Order Does Not Matter for Speech Recognition}
\name{Vineel Pratap\sthanks{Equal contribution.}, Qiantong Xu, Tatiana Likhomanenko\sthanks{Currently at Apple.}, Gabriel Synnaeve, Ronan Collobert$^*$}
\address{Facebook AI Research}
\begin{document}
\maketitle
\begin{abstract}
In this paper, we study training of automatic speech recognition system in a weakly supervised setting where the order of words in transcript labels of the audio training data is not known. We train a word-level acoustic model which aggregates the distribution of all output frames using $LogSumExp$ operation and uses a cross-entropy loss to match with the ground-truth words distribution. Using the pseudo-labels generated from this model on the training set, we then train a letter-based acoustic model using Connectionist Temporal Classification loss. Our system achieves $2.3\%$/$4.6\%$ on \textit{test-clean}/\textit{test-other} subsets of LibriSpeech, which closely matches with the supervised baseline{\textquotesingle}s performance.

\end{abstract}


\section{Introduction}
Transcribing speech, and generally labeling data, is an expensive process. 
Thus, it is relevant to study what level of supervision is needed in the first place. 
Indeed, sparse, crude annotations come cheaper, and can even sometimes be mined in the wild. 
Modern acoustic models (AMs) are able to classify so well that, we believe and particularly demonstrate in this paper, they can recover at least word order, and probably much more.
This ability revolves around the fact that self-training (pseudo-labeling)~\cite{lee2013pseudo} works even with models that are far from convergence and have only a weak performance~\cite{likhomanenko2020slimipl}. 
Said differently, it is possible for the model to improve by training with \textit{very noisy} labels. 
What other type of noise in the labels distribution can a model overcome, by recouping co-occurrences through statistic of a large enough training set?

We investigate if an automatic speech recognition (ASR) model can be trained with the sole annotations being the distribution of labels (bag of words), with a restricted (words) vocabulary.
It turns out that combining this kind of model with self-training reaches the same performance as a fully supervised equivalent model.
Thus, the word order has no importance (the title of this paper), as it can be easily recovered by the model and self-training: without a language model, with a simple training scheme, at least with enough data (960 hours of the LibriSpeech~\cite{povey2015} benchmark).

\section{Related Work}
{\bf Weakly supervised learning} can be formulated in various ways depending on how the weakly supervised labels are defined. 
Several works study weakly supervised training with video data: either contextual metadata~\cite{singh2020large} or subtitles\footnote{Authors design an algorithm to automatically recognize subtitles (can be viewed as "noisy" labels).} presented as a part of a video frame~\cite{cheng2020weakly}.
Both works demonstrate that weakly supervised training in combination with supervised training improves ASR over standalone supervised training.
In contrast, we use another source of weak supervision. 
\textit{Bag-of-words} labels for each sample (hard labels) has been considered in~\cite{palaz2016} to classify words, while we use words distribution (soft labels). 
\cite{kamper2017visually} also considers soft labels as we do, but these soft labels do not form a distribution over the words: image-to-words visual classifier tags images with soft textual multi-labels.
Being similar to us in the weak supervision formulation, both~\cite{palaz2016,kamper2017visually} are solving a (semantic) keyword spotting task, while we focus on more general ASR task.
Related to the \textit{bag-of-words} approaches, \cite{richard2018action} introduces a system for weakly supervised temporal action segmentation and labeling given only unordered action sets for video.

{\bf Semi-supervised learning} has been actively studied over last years in speech recognition community, primarily in the context of self-training, or pseudo-labeling, \cite{kahn2020self,chen2020semi,xu2020iterative,park2020improved,likhomanenko2020slimipl,higuchi2021momentum,zhang2020pushing,moritz2021semi}. 
All these approaches combine both labeled and unlabeled data with pseudo-labels (PLs), generated in one way or another. We use PLs in teacher-student manner to further improve weakly supervised model switching from word tokens to letter tokens. 

{\bf Unsupervised pre-training} on unlabeled data has also been used  recently to improve ASR task with further fine-tuning on labeled data~\cite{baevski2020wav2vec,hsu2021hubert,ravanelli2020multi,liu2021tera,sadhu2021wav2vec} or alternating supervised and unsupervised losses optimization~\cite{talnikar2021joint}. Others focused on learning representations from unlabeled data which are useful for phoneme classification and speaker verification~\cite{chung2019unsupervised,liu2020non,liu2021tera} (labeled data are still used to train a classifier on top of learned representations). 
Compared to these efforts we use only weak supervision.

{\bf Unsupervised learning} for ASR is an ongoing research on methods with no supervision at all: they either learn how to align unlabeled text and unlabeled audio~\cite{yeh2018unsupervised} or use adversarial training~\cite{liu2018completely,chen2019completely,baevski2021unsupervised}.

\begin{figure}[t!]
    \centering
    \advance\leftskip1cm
    \includegraphics[width=\linewidth]{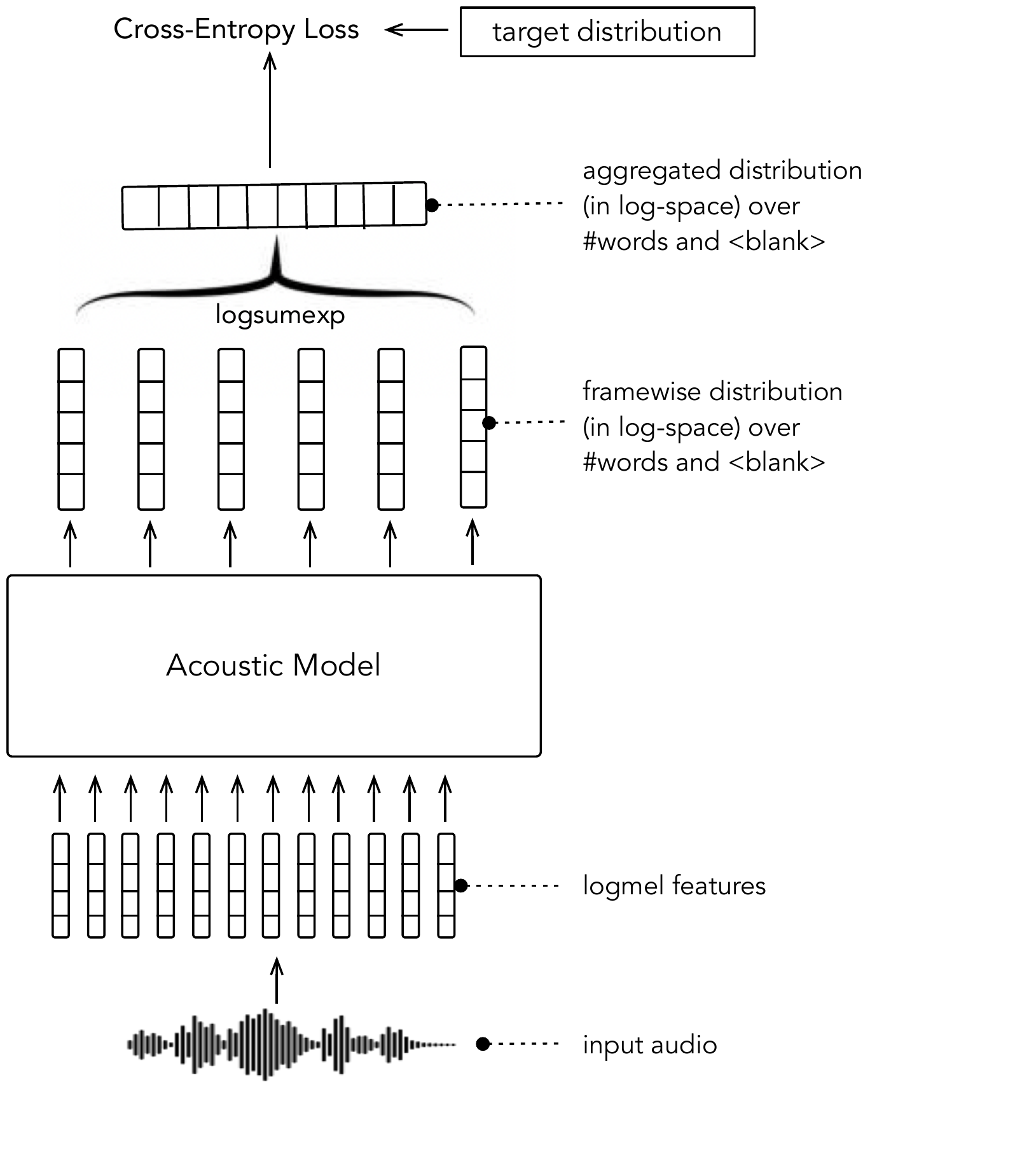}
    \vspace{-1cm}
    \caption{Weakly supervised ASR training pipeline.}
    \label{fig:system}
\end{figure}

\section{Method}
\subsection{Overview}
We consider the problem of performing ASR in a weakly supervised manner where the order of words in the transcript labels of training data are not known. Traditionally, in a supervised setting for ASR, we are given a training set $\cup_{i=1}^{N} \{x_i, y_i\}$ comprising of $N$ samples, where  $x_i$ is an input audio sequence and $y_i$ is a sequence of words. In the weakly supervised setting that we consider, $y_i \in \mathbb{R}^{|V|+1}$ is a probability distribution over the word vocabulary $V$ and a  \textit{<blank>} token which models all ``garbage'' frames that could occur between words. All out-of-vocabulary (OOV) words are mapped to a special \textit{<unk>} word in the vocabulary. An overview of the full training pipeline used in our work is presented in Algorithm~\ref{algo:ipl}.

\begin{algorithm}[h]

\small{
\KwData{Audio samples $\{x_i\}$ and their bag-of-words labels $\{y_i\}$.}
\KwResult{Acoustic model $g_\theta$.}
1. Train a word-based weakly supervised model $q_\theta$ on $\{x_i, y_i\}$ with cross-entropy loss, Eq.(\ref{eq:loss}), until convergence;

2. Apply greedy decoding on $\{x_i\}$ using $q_\theta$ to generate transcriptions $\{\tilde y_i\}$ as PLs, where $\tilde y_i = \mathrm{argmax}_y  q_\theta ( y | x_i )$;

3. Replace $\textit{<unk>}$ word in PLs using an $n$-gram LM beam-search decoding with a constraint that the words to be replaced should belong to the corresponding transcription;   

4. Train a letter-based ASR model $g_\theta$ with CTC loss using the PLs generated from Step 3. 

 \caption{Full training pipeline.}
\label{algo:ipl}
}
\end{algorithm}


\subsection{Target labels as probability distribution}

Let a target transcript in a supervised ASR setting is ``$w_0$ $w_1$ $w_2$ $w_1$'', where $w_0, w_1$ are the words in the vocabulary~$V$ and $w_2$ is an OOV word.  In our weakly supervised setting, we convert this into probability distribution by taking their count and normalizing the count by total number of words in the transcript. For example, the target for the above words sequence becomes $p=$\{$w_0$: $0.25$, $w_1$: $0.5$, $\textit{<unk>}$: $0.25$\}. Further, we also introduce a new hyperparameter $\alpha$ which is the prior probability on the \textit{<blank>} word and then re-normalize the probabilities of the words such that they sum to 1. For example, with $\alpha = 0.5$, the target label becomes  $p=$\{$w_0$: $0.125$, $w_1$: $0.25$, $\textit{<unk>}$: $0.125$, $\textit{<blank>}$ : $0.5$\}.      

\subsection{Weakly supervised word-level model training}\label{sec:am}
\label{weak_sup_am}
Fig.~\ref{fig:system} gives an overview of weakly supervised training. We use 80-dimensional log-mel spectrograms as input features. The AM architecture closely follows~\cite{likhomanenko2020rethinking}: the encoder is composed of a convolutional frontend (1-D convolution with kernel-width 7 and stride 3 followed by GLU activation) followed by 36 4-heads Transformer blocks~\cite{vaswani2017attention} with  relative positional embedding. The self-attention dimension is 384 and the feed-forward network (FFN) dimension is 3072 in each Transformer block. The output of the encoder is followed by a linear layer to the output classes. For all Transformer layers, we use dropout on the self-attention and on the FFN, and layer drop~\cite{fan2020}, dropping entire layers at the FFN level.  

We apply $LogSoftmax$ operation on each output frame to produce a probability distribution (in log-space) over output classes (vocabulary, $V$ + \textit{<blank>}). Then, all the output frames~$o$ are aggregated into a single probability distribution~$q$ by applying $LogSumExp$ operation with normalization by the number of output frames~$T$, see Eq.(\ref{eq:q}).
The loss function $\ll$ is the cross-entropy loss between predicted output distribution $q$ and the target distribution~$p$, see Eq.(\ref{eq:loss}).
\begin{equation}\label{eq:q}
q = LogSumExp\:(o_1, o_2, ... o_T) - log \: T  \\
\end{equation}
\begin{equation}\label{eq:loss}
\ll = - \sum_{i=1}^{|V|+1} p_i \;  log\: q_i
\end{equation}

\subsection{Inference using greedy decoding}
To perform inference using the weakly supervised model, we adopt a simple greedy decoding strategy. This is very similar to the greedy decoding procedure for Connectionist Temporal Classification  (CTC)~\cite{graves2006connectionist}. First, we pass the input utterance through the AM and get framewise emissions over the vocabulary $V$ and $\textit{<blank>}$ token. We then consider the word with max score at each time step, collapse the repeated tokens and remove $\textit{<blank>}$ token to get the final prediction. 

\subsection{From word-based to letter-based acoustic model}

The number of words in the vocabulary $V$ impacts both the training convergence and word error rate (WER) performance of our word-level AM. First, rare words appear only a couple of times in the training dataset\footnote{In the LibriSpeech training corpus, the 10,000 most frequent word appears 51 times.}, and thus the word-level AMs will not be able to generalize well on these words. Techniques to alleviate this issue have been explored in~\cite{settle2019words,collobert:2020}. However, we found in practice that with $|V| > 15,000$ the weakly supervised problem~(\ref{eq:loss}) is becoming very difficult, and with larger vocabularies the approach would fail to converge. We thus propose here to train a \emph{word}-based AM on a limited vocabulary in a \emph{weakly supervised} fashion, and then to "distill" its knowledge into a \emph{letter}-based AM, which has the potential to address any word in the dictionary.
Distillation is performed by running the word-based AM inference over the training utterances. We then train a letter-based AM on the corresponding generated PLs, via a regular CTC approach. We use the same Transformer-based encoder consisting of 270M parameters from \cite{likhomanenko2020rethinking} for the AM.

\subsection{Uncovering \textit{<unk>} in pseudo-labels (PLs)}
\label{sec:unk}
As the word-level AM is trained on a limited vocabulary $V$, corresponding PLs generated by this model may contain \textit{<unk>} words. While we can train the letter-based model by simply removing \textit{<unk>}, we will show that refining PLs by uncovering unknown words leads to better performing letter-based AMs. For that matter, we consider a language model (LM) $p^{LM}(\cdot)$ trained on a separate text-only training corpus\footnote{To ensure acoustic training transcriptions word order is not used, the language model training data should not include any acoustic transcriptions.}, with a large vocabulary $V^{LM}$. Then, considering a PL sequence $\pi = \{\pi_1,\, \pi_2,\, \dots,\,\pi_L\}$
with $L$ words, we denote ${\cal U}(\pi)$ the set of positions where an \textit{<unk>} was produced. We then aim at replacing these unknown words in the PL sequence by finding the most appropriate words according to the LM likelihood:
\begin{equation}
\label{eq:uncoverunk}
    \max_{\forall i \in {\cal U}(\pi),\, \pi_i \in V^{LM}} p^{LM}(\pi)
\end{equation}
For efficiency, maximizing this likelihood is performed with a beam-search procedure. It is possible to further constrain this search for unknown words replacement, by enforcing new words $\pi_i$ to belong to the original bag-of-word acoustic transcription, instead of the full vocabulary $\pi_i \in V^{LM}$ in $(\ref{eq:uncoverunk})$.




\section{Experiments and Results}
\label{sec:results}

All the models are trained using wav2letter++ \cite{pratap2018} framework. The experiments are run on Nvidia Volta 32GB GPUs and we use 16 GPUs for weakly supervised experiments and 64 GPUs for running CTC experiments. We use SpecAugment~\cite{park_2019} as the data augmentation to augment the input data for both weakly supervised, CTC model training. We use LibriSpeech~\cite{povey2015} dataset for our study which consists of 960h of read speech and report numbers on the standard dev/test sets. We use 5-gram LM for beam-search decoding of models trained with CTC and Transformer LM for rescoring the top hypothesis from beam-search to further improve WER performance. All the LMs are trained on the official LM training data provided with LibriSpeech. For details on beam-search decoding and Transformer LM rescoring used in our work, we refer the reader to \cite{synnaeve2019end}. 
\vspace{-0.1cm}
\subsection{Tuning \textit{<blank>} prior probability, \texorpdfstring{$\boldsymbol{\alpha}$}{alpha}}
\begin{figure}[h!]
\centering
\begin{tikzpicture}
    \begin{axis}[
        xlabel=\textit{<blank>} prior probability $\alpha$,
        ylabel=Greedy WER, 
        axis x line*=bottom,
        axis y line*=left,
        width=\linewidth,
        height=6cm,
        ytick={0,25,50,75,100},]
      
     \addplot[mark=*,RoyalBlue] plot coordinates {
        (0, 100)
        (0.1, 91.8)
        (0.2, 82.22)
        (0.3,70.705)
        (0.4,51.22)
        (0.5,34.0557)
        (0.6,19.61)
        (0.7,11.55)
        (0.8,9.0)
        (0.9,7.8)
    };
    \addlegendentry{dev-clean}
    
    \addplot[mark=*,RedViolet] plot coordinates {
        (0, 100)
        (0.1, 92.5)
        (0.2,  84.96)
        (0.3,74.18)
        (0.4,55.77)
        (0.5,38.63)
        (0.6,24.67)
        (0.7,15.89)
        (0.8,13.57)
        (0.9,10.75)
    };
   \addlegendentry{dev-other}    
\end{axis}
\end{tikzpicture}
\caption{Dependence between WER (no LM) and $\alpha$ hyperparameter on \textit{dev-clean} and \textit{dev-other} subsets of LibriSpeech.} \label{fig:alphatune}
\end{figure}
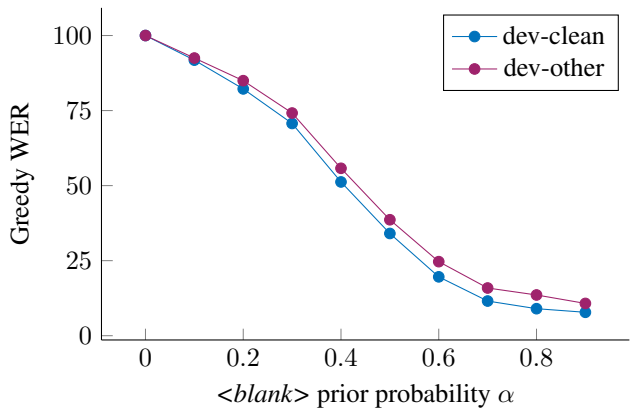

\vspace{-0.1cm}
We performed a grid search over $\textit{<blank>}$ token prior  probability, $\alpha$ values from $0$ to $0.9$ in steps of $0.1$. We use a vocabulary size of 10K, which consists of top 10K words from the training set sorted by their frequency. Fig.~\ref{fig:alphatune} shows the word error rate (WER) with greedy decoding (no LM) on clean/other dev subsets of LibriSpeech with varying  hyperparameter $\alpha$: the best performance is achieved for $\alpha = 0.9$. It is interesting to note that ${\alpha = 0.9}$ roughly corresponds to predicting each word for one output frame and predicting $\textit{<blank>}$ for all other frames. This follows from the fact that the AM outputs an average of $33.33$ frames per second when using stride $3$ in the convolution layer and $10$ms hop length for Short Time Fourier Transform (STFT) computation. Also, the samples in LibriSpeech have an average speaking rate of 2.7 words per second. Thus, predicting each word for one output frame and \textit{<blank>} for all other output frames amounts to $\alpha = 1.0 - 2.7/33.33 = 0.91$.        
 
\vspace{-0.5cm}
\subsection{Weakly Supervised ASR} 
 For weakly supervised training, we use top 10k words (based on the frequency) from the training set as the vocabulary. Table 1 shows results of weakly supervised ASR model which achieves $8.2\%/11.3\%$ WER (no LM) on the clean/other test subsets, respectively. Since the model is trained only to infer 10k words, it is not a fair comparison to compare the WER numbers to supervised baseline model. 
 
\begin{table}[h!]
\caption{WER comparison for supervised and weakly supervised models on LibriSpeech. \label{tab:ls_wer}}

\centering
\resizebox{\linewidth}{!}{
\begin{tabular}{@{}cccccc@{}}
\toprule
\multirow{2}{*}{\textbf{Method}} &
\multirow{2}{*}{\textbf{LM}} &
\multicolumn{2}{c}{\textbf{Dev WER}} & 
\multicolumn{2}{c}{\textbf{Test WER}} \\
\cmidrule(lr){3-4} \cmidrule(lr){5-6}  
& & clean & other & clean & other \\
\midrule
\multirow{2}{*}{Supervised \cite{zhang2020pushing} (SOTA) } & - & 1.9 & 4.4 & 2.1 & 4.3 \\
 & Transformer  & & & 1.9 & 3.9 \\
\midrule
\midrule
\multirow{3}{*}{Supervised (baseline)}  & - & 2.5 & 5.9 & 2.7 & 6.1 \\
& word 5-gram  & 1.9 & 4.7 & 2.4 & 5.3 \\
& Transformer  & 1.6 & 4.0 & 2.1 & 4.5 \\
\midrule
\midrule
\begin{tabular}{@{}c@{}}Weakly supervised \\ (word-based)\end{tabular}     & - & 7.8 & 10.7 & 8.2 & 11.3 \\
\midrule
 \multirow{3}{*}{$\quad +$ PL (letter-based)} & - & 2.9 & 6.4 & 3.0 & 6.5 \\
 & word 5-gram  & 2.3 & 5.2 & 2.6 & 5.5  \\
& Transformer  & 1.9 & 4.3 & 2.3  & 4.6 \\
\bottomrule
\end{tabular}
}
\vspace{-7pt}
\end{table}

\begin{table}[t!]
\caption{WER for PLs generated with $3$ different strategies to deal with $\textit{<unk>}$: (1) remove $\textit{<unk>}$ word in PLs; (2) and (3) replace $\textit{<unk>}$ in PLs with beam-search procedure using 5-gram LM over full train vocabulary or transcript vocabulary of the corresponding sample.}
\label{tab:pl_wer}
\centering
\resizebox{\linewidth}{!}{
\begin{tabular}{@{}lccc@{}}
\toprule
\multirow{2}{*}{\textbf{Strategy}} &
\multirow{2}{*}{\textbf{Train WER}} &
\multicolumn{2}{c}{\textbf{Dev WER}} \\
\cmidrule(lr){3-4}
& & clean & other \\
\midrule
1. Remove \textit{<unk>}  & 6.2 &  7.5 & 9.9  \\
2. Replace \textit{<unk>};  train vocab.  & 3.7 & 5.1 & 8.7  \\
3. Replace \textit{<unk>}; transcript vocab.  & 1.6 & 2.9 & 6.4 \\
\bottomrule

\end{tabular}
}
\vspace{-7pt}
\end{table}

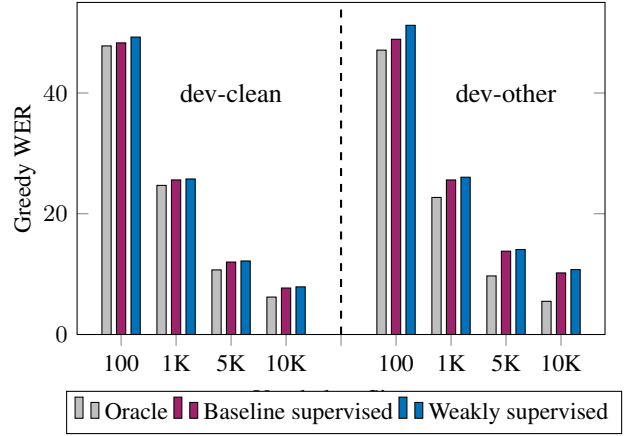
\begin{figure}[t!]
\centering
\begin{tikzpicture}
    \begin{axis}[
        xlabel=Vocabulary Size,
        ylabel=Greedy WER, 
        ylabel style={yshift=-0.4em, font=\small},
        y tick label style={ font=\small},
        xlabel style={xshift=-0.4em, font=\small},
        x tick label style={ font=\small},
        xtick pos=left,
        ybar  ,
        width=\linewidth,
        height=6cm,
        bar width = 3.5pt,
        xticklabels={100, 1K, 5K, 10K,,100, 1K, 5K, 10K},
        ymin=0,ymax=55,
        xtick={0,1,2,3,4,5,6,7,8},
        legend cell align={left},
        legend columns=-1,
        legend style={font=\small,at={(axis cs:-1,-16.5)},anchor=south west}
        ]
 
   \addplot[fill=lightgray,bar group size={0}{3}] coordinates {
        (0, 47.8)
        (1, 24.7)
        (2, 10.7)
        (3, 6.2)
    };
    \addplot[fill=RedViolet,bar group size={1}{3}] coordinates {
        (0, 48.3)
        (1, 25.6)
        (2, 12.0)
        (3, 7.7)
    };
    
    \addplot[fill=RoyalBlue,bar group size={2}{3}] coordinates {
        (0, 49.26)
        (1, 25.76)
        (2, 12.18)
        (3, 7.89)
    };
      \addplot[fill=lightgray, bar group size={0}{3}] coordinates {
        (5, 47.1)
        (6, 22.7)
        (7, 9.7)
        (8, 5.5)
    };
     \addplot[fill=RedViolet,bar group size={1}{3}] coordinates {
        (5, 48.9)
        (6, 25.6)
        (7, 13.8)
        (8, 10.2)
    };
   
   \addplot[fill=RoyalBlue,bar group size={2}{3}] coordinates {
        (5, 51.21)
        (6, 26.05)
        (7, 14.08)
        (8, 10.75)
    };
    
  \path
    (axis cs:3, \pgfkeysvalueof{/pgfplots/ymin})
    -- coordinate (tmpmin)
    (axis cs:5, \pgfkeysvalueof{/pgfplots/ymin})
    (axis cs:3, \pgfkeysvalueof{/pgfplots/ymax})
    -- coordinate (tmpmax)
    (axis cs:5, \pgfkeysvalueof{/pgfplots/ymax})
  ;
  \draw[thick, dashed] (tmpmin) -- (tmpmax);
   \node at (axis cs:2,40) {dev-clean}; 
    \node at (axis cs:7,40) {dev-other}; 
\legend{Oracle , Baseline supervised, Weakly supervised  }  

\end{axis}
\end{tikzpicture}
\caption{WER comparison for different constrained lexicon sizes on dev sets }\label{fig:wer_comp}
\end{figure}

To get a sense of how good the weakly supervised model is performing, we have trained 4 different models with vocabulary sizes of 100, 1K, 5K and 10K words with $\alpha=0.9$. We compare their performance with WER of our supervised baseline model where the output vocabulary is restricted to the same vocabulary. We also measure the WER of oracle model which always outputs correct predictions and is restricted to the output vocabulary, serving as a lower bound for the WER. Fig.~\ref{fig:wer_comp} shows the WER comparison for different vocabulary sizes on the dev sets. It can be seen that the the weakly supervised model can achieve a performance competitive with the supervised model (after restricting the output vocabulary). 
 
\begin{figure}[t!]
    \centering
    \includegraphics[width=\linewidth]{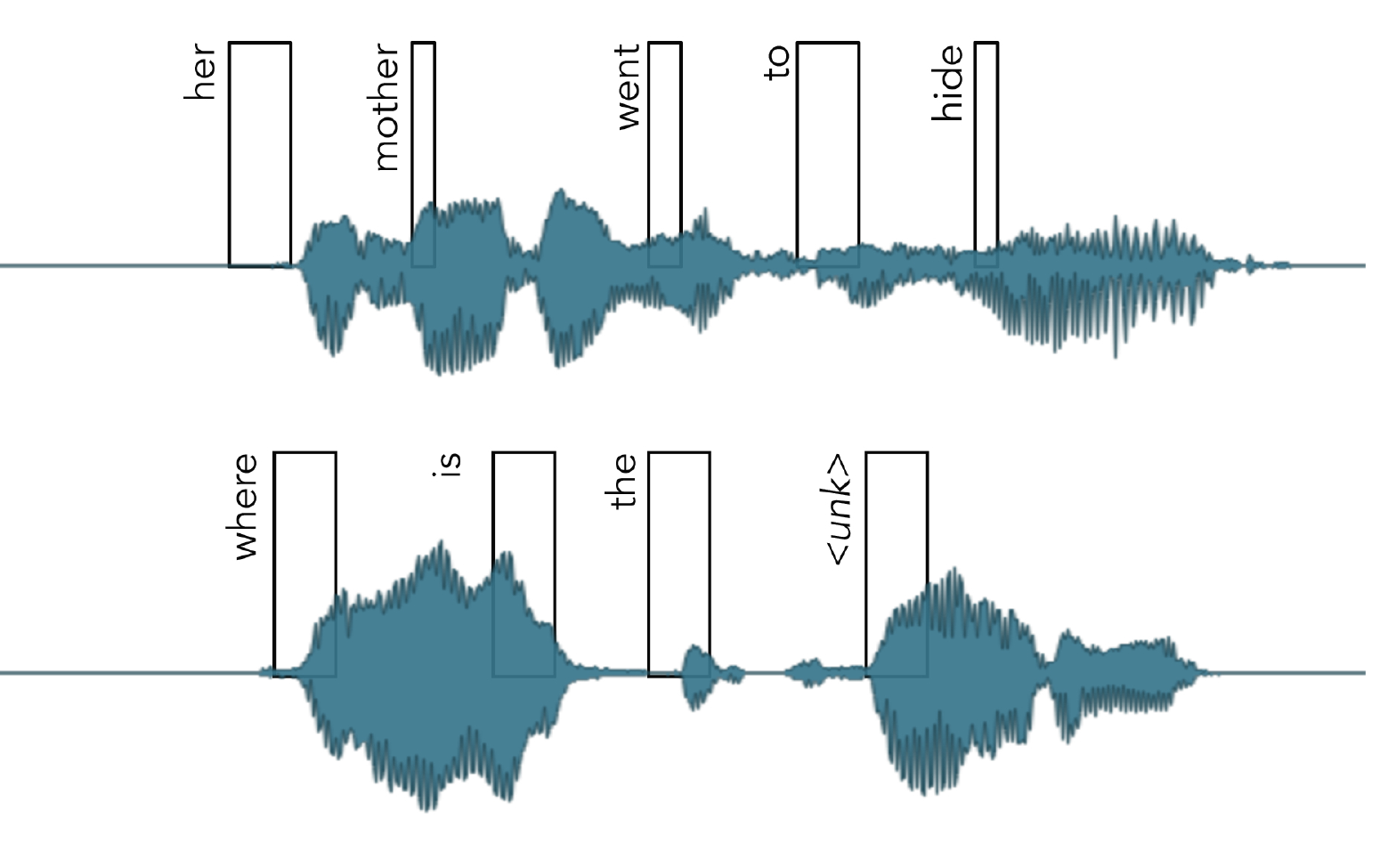}
    \vspace{-20pt}
    \caption{Word localisation for the weakly supervised model.}
    \label{fig:segment}
    \vspace{-10pt}
\end{figure}

In order to understand if the model is able to localise word predictions~\cite{palaz2016}, we perform inference on the audio file taking the output tokens with highest score at each output frame. We map the output frames corresponding to the inferred sequence back to input audio to see where the words are produced. A few examples are shown in Fig.~\ref{fig:segment}. It can be seen the model tends to output a word at the onset of the word in the audio. 
\vspace{-0.3cm}
\subsection{Pseudo-Labeling} 
We use the weakly supervised model trained on 10K vocabulary to generate PLs on training set. As we discussed in Section \ref{sec:unk}, we consider 3 possible ways to deal with $\textit{<unk>}$ word in PLs and the WER on training set is shown in Table \ref{tab:pl_wer}. We can see that replacing $\textit{<unk>}$ in the PLs with beam-search using 5-gram LM and restricting vocabulary to the original bag-of-word acoustic transcriptions gives the best performance. We train letter-based CTC models on these PLs and the performance on dev sets is shown in Table~\ref{tab:pl_wer}: the better WER on the training set translates to the better WER on dev sets.

And finally, we perform beam-search decoding using 5-gram LM for the best model which uses PLs from strategy~(3): it gives a WER of $2.6\%$/$5.5\%$ on clean/other test sets, respectively. Rescoring with a Transformer LM further improves the WER to $2.3\%$/$4.6\%$ on clean/other test sets and is competitive with the supervised baseline.  
\vspace{-0.2cm}
\section{Conclusion}
\vspace{-0.2cm}
We demonstrated that speech recognition models can be developed without any knowledge of the words order or their counts in the transcript, and words distribution is enough. Moreover, weakly supervised learning can be successfully combined together with pseudo-labeling to achieve the same performance as supervised learning. This result shows that state-of-the-art speech recognition models can be developed with much less supervision than what is traditionally being used. 


\clearpage

\bibliographystyle{IEEEbib}

\footnotesize{\bibliography{references}}

\begin{thebibliography}{10}

\bibitem{lee2013pseudo}
Dong-Hyun Lee,
\newblock ``Pseudo-label: The simple and efficient semi-supervised learning
  method for deep neural networks,''
\newblock in {\em Workshop on challenges in representation learning, ICML},
  2013, vol.~3.

\bibitem{likhomanenko2020slimipl}
Tatiana Likhomanenko, Qiantong Xu, Jacob Kahn, Gabriel Synnaeve, and Ronan
  Collobert,
\newblock ``slimipl: Language-model-free iterative pseudo-labeling,''
\newblock in {\em Interspeech}, 2021.

\bibitem{povey2015}
V.~{Panayotov}, G.~{Chen}, D.~{Povey}, and S.~{Khudanpur},
\newblock ``Librispeech: An asr corpus based on public domain audio books,''
\newblock in {\em 2015 IEEE International Conference on Acoustics, Speech and
  Signal Processing (ICASSP)}, 2015, pp. 5206--5210.

\bibitem{singh2020large}
Kritika Singh, Vimal Manohar, Alex Xiao, Sergey Edunov, Ross Girshick, Vitaliy
  Liptchinsky, Christian Fuegen, Yatharth Saraf, Geoffrey Zweig, and
  Abdelrahman Mohamed,
\newblock ``Large scale weakly and semi-supervised learning for low-resource
  video asr,''
\newblock in {\em Interspeech}, 2020.

\bibitem{cheng2020weakly}
Mengli Cheng, Chengyu Wang, Xu~Hu, Jun Huang, and Xiaobo Wang,
\newblock ``Weakly supervised construction of asr systems with massive video
  data,''
\newblock {\em arXiv preprint arXiv:2008.01300}, 2020.

\bibitem{palaz2016}
Dimitri Palaz, Gabriel Synnaeve, and Ronan Collobert,
\newblock ``Jointly learning to locate and classify words using convolutional
  networks,''
\newblock in {\em Interspeech}, 2016, pp. 2741--2745.

\bibitem{kamper2017visually}
Herman Kamper, Shane Settle, Gregory Shakhnarovich, and Karen Livescu,
\newblock ``Visually grounded learning of keyword prediction from untranscribed
  speech,''
\newblock in {\em Interspeech}, 2017.

\bibitem{richard2018action}
Alexander Richard, Hilde Kuehne, and Juergen Gall,
\newblock ``Action sets: Weakly supervised action segmentation without ordering
  constraints,''
\newblock in {\em Proceedings of the IEEE conference on Computer Vision and
  Pattern Recognition}, 2018, pp. 5987--5996.

\bibitem{kahn2020self}
Jacob Kahn, Ann Lee, and Awni Hannun,
\newblock ``Self-training for end-to-end speech recognition,''
\newblock in {\em ICASSP 2020-2020 IEEE International Conference on Acoustics,
  Speech and Signal Processing (ICASSP)}. IEEE, 2020, pp. 7084--7088.

\bibitem{chen2020semi}
Yang Chen, Weiran Wang, and Chao Wang,
\newblock ``Semi-supervised asr by end-to-end self-training,''
\newblock {\em Proc. Interspeech 2020}, pp. 2787--2791, 2020.

\bibitem{xu2020iterative}
Qiantong Xu, Tatiana Likhomanenko, Jacob Kahn, Awni Hannun, Gabriel Synnaeve,
  and Ronan Collobert,
\newblock ``Iterative pseudo-labeling for speech recognition,''
\newblock in {\em Interspeech}, 2020, pp. 1006--1010.

\bibitem{park2020improved}
Daniel~S Park, Yu~Zhang, Ye~Jia, Wei Han, Chung-Cheng Chiu, Bo~Li, Yonghui Wu,
  and Quoc~V Le,
\newblock ``Improved noisy student training for automatic speech recognition,''
\newblock in {\em Interspeech}, 2020, pp. 2817--2821.

\bibitem{higuchi2021momentum}
Yosuke Higuchi, Niko Moritz, Jonathan~Le Roux, and Takaaki Hori,
\newblock ``Momentum pseudo-labeling for semi-supervised speech recognition,''
\newblock in {\em Interspeech}, 2021.

\bibitem{zhang2020pushing}
Yu~Zhang, James Qin, Daniel~S. Park, Wei Han, Chung-Cheng Chiu, Ruoming Pang,
  Quoc~V. Le, and Yonghui Wu,
\newblock ``Pushing the limits of semi-supervised learning for automatic speech
  recognition,'' 2020.

\bibitem{moritz2021semi}
Niko Moritz, Takaaki Hori, and Jonathan Le~Roux,
\newblock ``Semi-supervised speech recognition via graph-based temporal
  classification,''
\newblock in {\em ICASSP 2021-2021 IEEE International Conference on Acoustics,
  Speech and Signal Processing (ICASSP)}. IEEE, 2021, pp. 6548--6552.

\bibitem{baevski2020wav2vec}
Alexei Baevski, Yuhao Zhou, Abdelrahman Mohamed, and Michael Auli,
\newblock ``wav2vec 2.0: A framework for self-supervised learning of speech
  representations,''
\newblock {\em Advances in Neural Information Processing Systems}, vol. 33,
  2020.

\bibitem{hsu2021hubert}
Wei-Ning Hsu, Benjamin Bolte, Yao-Hung~Hubert Tsai, Kushal Lakhotia, Ruslan
  Salakhutdinov, and Abdelrahman Mohamed,
\newblock ``Hubert: Self-supervised speech representation learning by masked
  prediction of hidden units,''
\newblock {\em arXiv preprint arXiv:2106.07447}, 2021.

\bibitem{ravanelli2020multi}
Mirco Ravanelli, Jianyuan Zhong, Santiago Pascual, Pawel Swietojanski, Joao
  Monteiro, Jan Trmal, and Yoshua Bengio,
\newblock ``Multi-task self-supervised learning for robust speech
  recognition,''
\newblock in {\em ICASSP 2020-2020 IEEE International Conference on Acoustics,
  Speech and Signal Processing (ICASSP)}. IEEE, 2020, pp. 6989--6993.

\bibitem{liu2021tera}
Andy~T Liu, Shang-Wen Li, and Hung-yi Lee,
\newblock ``Tera: Self-supervised learning of transformer encoder
  representation for speech,''
\newblock {\em IEEE/ACM Transactions on Audio, Speech, and Language
  Processing}, vol. 29, pp. 2351--2366, 2021.

\bibitem{sadhu2021wav2vec}
Samik Sadhu, Di~He, Che-Wei Huang, Sri~Harish Mallidi, Minhua Wu, Ariya
  Rastrow, Andreas Stolcke, Jasha Droppo, and Roland Maas,
\newblock ``Wav2vec-c: A self-supervised model for speech representation
  learning,''
\newblock {\em arXiv preprint arXiv:2103.08393}, 2021.

\bibitem{talnikar2021joint}
Chaitanya Talnikar, Tatiana Likhomanenko, Ronan Collobert, and Gabriel
  Synnaeve,
\newblock ``Joint masked cpc and ctc training for asr,''
\newblock in {\em ICASSP 2021-2021 IEEE International Conference on Acoustics,
  Speech and Signal Processing (ICASSP)}. IEEE, 2021, pp. 3045--3049.

\bibitem{chung2019unsupervised}
Yu-An Chung, Wei-Ning Hsu, Hao Tang, and James Glass,
\newblock ``An unsupervised autoregressive model for speech representation
  learning,''
\newblock in {\em Interspeech}, 2019.

\bibitem{liu2020non}
Alexander~H Liu, Yu-An Chung, and James Glass,
\newblock ``Non-autoregressive predictive coding for learning speech
  representations from local dependencies,''
\newblock {\em arXiv preprint arXiv:2011.00406}, 2020.

\bibitem{yeh2018unsupervised}
Chih-Kuan Yeh, Jianshu Chen, Chengzhu Yu, and Dong Yu,
\newblock ``Unsupervised speech recognition via segmental empirical output
  distribution matching,''
\newblock in {\em International Conference on Learning Representations}, 2019.

\bibitem{liu2018completely}
Da-Rong Liu, Kuan-Yu Chen, Hung-yi Lee, and Lin-shan Lee,
\newblock ``Completely unsupervised phoneme recognition by adversarially
  learning mapping relationships from audio embeddings,''
\newblock in {\em Interspeech}, 2018.

\bibitem{chen2019completely}
Kuan-Yu Chen, Che-Ping Tsai, Da-Rong Liu, Hung-Yi Lee, and Lin-shan Lee,
\newblock ``Completely unsupervised phoneme recognition by a generative
  adversarial network harmonized with iteratively refined hidden markov
  models.,''
\newblock in {\em Interspeech}, 2019, pp. 1856--1860.

\bibitem{baevski2021unsupervised}
Alexei Baevski, Wei-Ning Hsu, Alexis Conneau, and Michael Auli,
\newblock ``Unsupervised speech recognition,''
\newblock {\em arXiv preprint arXiv:2105.11084}, 2021.

\bibitem{likhomanenko2020rethinking}
Tatiana Likhomanenko, Qiantong Xu, Vineel Pratap, Paden Tomasello, Jacob Kahn,
  Gilad Avidov, Ronan Collobert, and Gabriel Synnaeve,
\newblock ``Rethinking evaluation in asr: Are our models robust enough?,''
\newblock in {\em Interspeech}, 2020.

\bibitem{vaswani2017attention}
Ashish Vaswani, Noam Shazeer, Niki Parmar, Jakob Uszkoreit, Llion Jones,
  Aidan~N. Gomez, Lukasz Kaiser, and Illia Polosukhin,
\newblock ``Attention is all you need,'' 2017.

\bibitem{fan2020}
Angela Fan, Edouard Grave, and Armand Joulin,
\newblock ``Reducing transformer depth on demand with structured dropout,''
\newblock in {\em 8th International Conference on Learning Representations,
  {ICLR} 2020, Addis Ababa, Ethiopia, April 26-30, 2020}. 2020, OpenReview.net.

\bibitem{graves2006connectionist}
Alex Graves, Santiago Fern{\'a}ndez, Faustino Gomez, and J{\"u}rgen
  Schmidhuber,
\newblock ``Connectionist temporal classification: labelling unsegmented
  sequence data with recurrent neural networks,''
\newblock in {\em Proceedings of the 23rd international conference on Machine
  learning}, 2006, pp. 369--376.

\bibitem{settle2019words}
Shane Settle, Kartik Audhkhasi, Karen Livescu, and Michael Picheny,
\newblock ``Acoustically grounded word embeddings for improved
  acoustics-to-word speech recognition,''
\newblock in {\em 2015 IEEE International Conference on Acoustics, Speech and
  Signal Processing (ICASSP)}. IEEE, 2019.

\bibitem{collobert:2020}
Ronan Collobert, Awni Hannun, and Gabriel Synnaeve,
\newblock ``Word-level speech recognition with a letter to word encoder,''
\newblock in {\em International Conference on Machine Learning, {ICML}}, 2020.

\bibitem{pratap2018}
V.~{Pratap}, A.~{Hannun}, Q.~{Xu}, J.~{Cai}, J.~{Kahn}, G.~{Synnaeve},
  V.~{Liptchinsky}, and R.~{Collobert},
\newblock ``Wav2letter++: A fast open-source speech recognition system,''
\newblock in {\em ICASSP 2019 - 2019 IEEE International Conference on
  Acoustics, Speech and Signal Processing (ICASSP)}, 2019, pp. 6460--6464.

\bibitem{park_2019}
Daniel~S. Park et~al.,
\newblock ``Specaugment: A simple data augmentation method for automatic speech
  recognition,''
\newblock {\em Interspeech 2019}, Sep 2019.

\bibitem{synnaeve2019end}
G.~Synnaeve et~al.,
\newblock ``End-to-end {ASR}: from {Supervised} to {Semi}-{Supervised}
  {Learning} with {Modern} {Architectures},''
\newblock {\em arXiv}, vol. abs/1911.08460, 2019.

\end{thebibliography}

\end{document}